\documentclass[twocolumn,prl]{revtex4}
\usepackage{graphicx}
\usepackage{color}
\usepackage{amsmath}

\newcommand{\be}{\begin{equation}}
\newcommand{\ee}{\end{equation}}
\newcommand{\bea}{\begin{eqnarray}}
\newcommand{\eea}{\end{eqnarray}}

\newcommand{\p}{\partial}

\newcommand{\la}{\langle}
\newcommand{\ra}{\rangle}
\newcommand{\lb}{\left[}
\newcommand{\rb}{\right]}
\newcommand{\lp}{\left(}
\newcommand{\rp}{\right)}

\renewcommand{\vec}[1]{{\bf #1}}

\renewcommand{\phi}{\varphi}
\renewcommand{\epsilon}{\varepsilon}

\begin{document}

\title{
Quasi-Relativistic Doppler Effect and Non-Reciprocal Plasmons in Graphene}

\author{Dan S. Borgnia, Trung V. Phan, Leonid S. Levitov}

\affiliation{Physics Department, Massachusetts Institute of Technology, 77 Massachusetts Avenue, Cambridge MA02139}

\begin{abstract}
Strong optical nonreciprocity at the nanoscale, 
relying on extreme one-way modes and backscattering suppression, can enable fundamentally new approaches in optoelectronics and plasmonics. 
Of special interest is achieving nonreciprocity in systems devoid of magnetic couplings. 
We describe a new approach based on the plasmonic Doppler effect which takes place for plasmons propagating in the presence of an electrical DC current. Large carrier drift velocities reachable in high-mobility electron systems, such as graphene, 
can enable strongly nonreciprocal or even fully one-way modes. Striking effects such as mode isolation and one-way transmission in DC-current-controlled Mach-Zehnder interferometers provide clear manifestations of plasmonic nonreciprocity. Phenomena such as plasmon resonance splitting into a doublet, induced by a DC current, afford new ways to generate and exploit unidirectionally propagating plasmon modes.
\end{abstract}

\maketitle
Novel one-way photonic and plasmonic modes are of keen fundamental interest and also
can harbor practical opportunities. 
Backscattering-free  modes are intrinsic to some materials: in topological materials backscattering is suppressed by  the topological protection effects,\cite{haldane_2008,wang_2008,khanikaev_2013} whereas in Dirac materials it is weak due to conserved chirality.\cite{katsnelson_2006,sepkhanov_2007,lu_2013}
A more flexible/radical approach, however, relies on introducing couplings that generate an asymmetry between counter-propagating modes in system bulk, $\omega_{-\vec k}\ne\omega_{\vec k}$, ultimately eliminating one of these modes. Such modes feature maximally robust uni-directional behavior known as `strong non-reciprocity'.\cite{jalas_2013}

Non-reciprocity in bulk materials can be realized by breaking time-reversal symmetry with the magnetic field. However, 
the magneto-optical effects---the main tool for achieving nonreciprocity---are unsuitable for nanoscale systems\cite{yu_2009,manipatruni_2009,fan_2011,kang_2011,kamal_2011,lira_2012}. 
Atomically thin materials such as graphene support novel deep-subwalength surface plasmon-polariton (SPP)  modes\cite{wunsch_2006,hwang_2007,polini_2009,hwang_2009}. 
The unique properties of these modes---the wide frequency range, the high degree of field confinement,  gate tunability, and low losses---make them a valuable addition to the nanophotonics toolbox\cite{bonaccorso_2010,koppens_2011,lu_2011,chen_2012,fei_2012}. It is thus of considerable interest to demonstrate nonreciprocity for such modes without relying on magnetic coupling. 
Nonlinear or time-dependent effects are helping tackle this challenge in traditional photonic materials\cite{yu_2009,manipatruni_2009,fan_2011,kang_2011,kamal_2011,lira_2012}. 
So far, however, it has proved challenging to demonstarte nonreciprocity in linear photonic or plasmonic systems without optical pumping.

\begin{figure}
\includegraphics[scale=0.06]{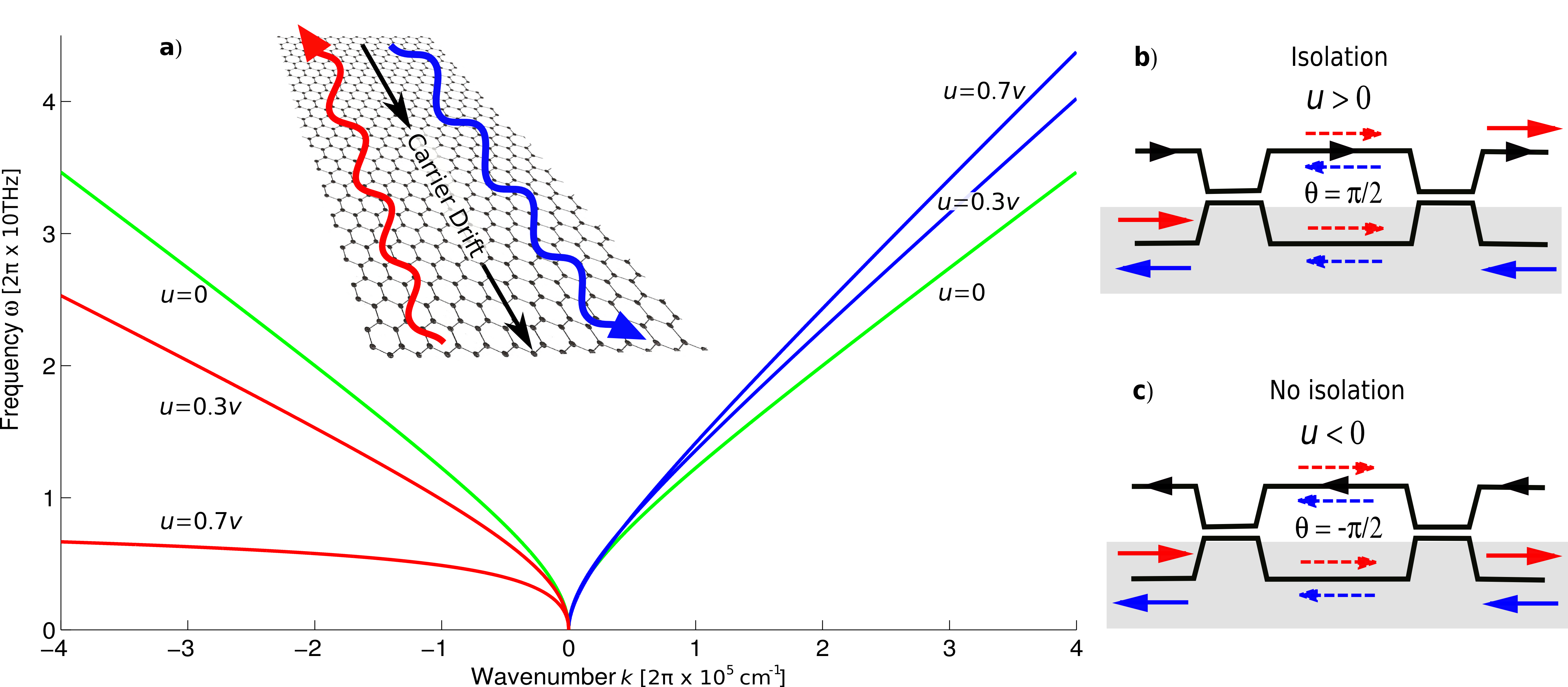} 
\caption{(a) Nonreciprocity induced by plasmonic Doppler effect in the presence of a DC current. Shown is the dispersion $\omega_{\vec k}$ for plasmons propagating 
along and against current for different drift velocity $u$ values (frequency blue-shifted and red-shifted,
 relative to the zero-current value, respectively). Doppler shift is first-order in $u/v$ for small $u$, and order-one at $u\sim v$, where $v=10^6 {\rm m/s}$ is the graphene carrier velocity. 
Plotted is the dispersion $\omega_{\vec k}$ given by Eq.(\ref{eq:dispersion_full}). Wavenumber and frequency units $k_*=2\pi \cdot 10^5\,{\rm cm}^{-1}$, $vk_*=2\pi\cdot 10\,{\rm THz}$, with $v=10^6\,{\rm m/s}$ and $k_*$ defined in Eq.(\ref{eq:delta_omega_2}), correspond to typical  parameters in the hydrodynamical regime, see text.
(b,c) Current-tunable Mach-Zehnder interferometer schematic. Plasmon modes couple via long-range Coulomb interaction in the `beamsplitter' regions. 
The interference phase, controlled by a DC current in one of the arms, can be tuned to achieve mode isolation and one-way transmission. 
} 
\label{fig1}
\end{figure}

Here we propose a very different approach based on the plasmonic Doppler effect  under an applied electrical DC current, wherein 
downstream (upstream) propagation results in a blue (red) frequency shift. 
Since time reversal symmetry is broken in the presence of a flow, neither nonlinear coupling nor pumping are required to produce nonreciprocity. 
The carrier drift velocity  in modern high-mobility 2DES can reach a substantial fraction of the Fermi velocity, leading to a strong Doppler effect. Measurements in graphene\cite{pop2010} report drift velocity values (in the saturation regime at strong fields) exceeding $3 \cdot 10^5\,{\rm m/s}$, which is a substantial fraction of carrier velocity in this material, $v=10^6\,{\rm m/s}$. The maximum drift velocity known, extracted from current density,\cite{balandin2012} is about $0.5v$. Similar values have been found elsewhere.\cite{perebeinos} 
These velocity values are orders of magnitude higher than maximum drift velocity values in metals.

The broad-band character of the Doppler effect as well as its strength in 2DES make it a suitable vehicle for achieving strong nonreciprocity. While our discussion will focus on graphene, we note that qualitative conclusions are similar for other high-mobility 2DES such as semiconductor quantum wells.\cite{theis_1980,chaplik_1985} Low carrier densities and high mobilities in these materials enable high drift velocities. As illustrated in Fig.\ref{fig1}, the Doppler effect can lead to strong nonreciprocity wherein SPP modes acquire a preferred propagation direction, $\omega_{-\vec k}\ne\omega_{\vec k}$, losing one of counter-propagating modes at large enough $u$.

\begin{figure}[t]
\includegraphics[scale=0.37]{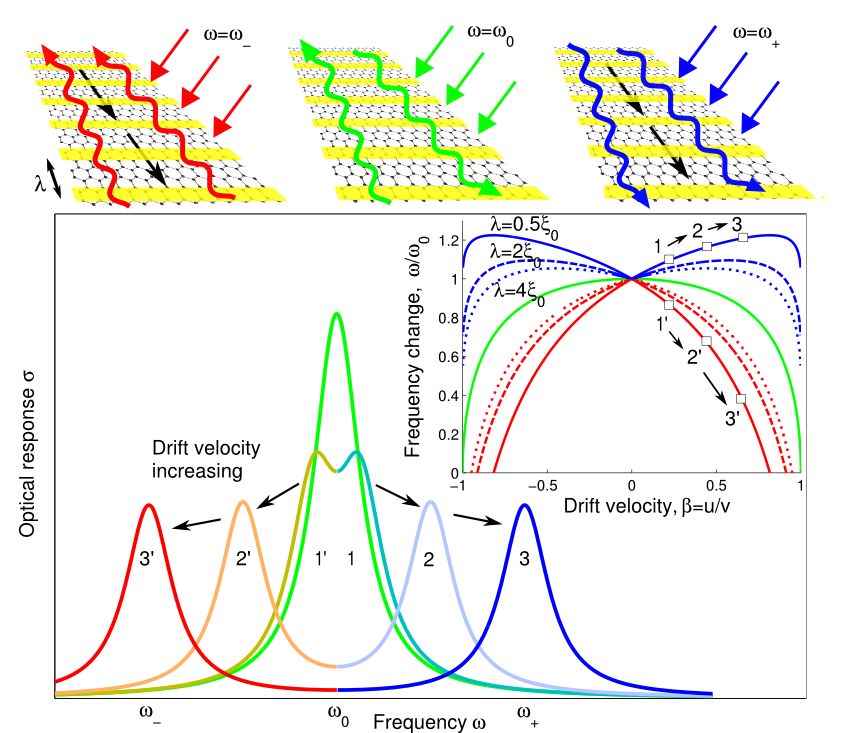}
\caption{Plasmon resonance in a patterned graphene ribbon undergoes splitting due to current-induced Doppler effect.
In the presence of a grating, 3D radiation can excite 2D modes with frequencies set by resonance condition $\omega_\pm(k)=\omega$, $k=\pm \frac{2\pi}{\lambda}$, with $\omega$ the 3D radiation frequency and $\lambda$ the grating periodicity. 
The resonance is blue-shifted (red-shifted) from zero-current value $\omega_0=\omega$ for plasmons propagating 
along (against) the flow.  Each of the two modes can be excited individually provided the flow-induced Doppler splitting, $\Delta=\omega_+ - \omega_-$, is larger than the resonance width.
} 
\label{fig2}
\end{figure}

One appealing way detect and probe  nonreciprocity in the Doppler effect regime employs current-tunable
Mach-Zehnder interferometers, wherein the interference phase is controlled by a DC current in one of the arms, see Fig.\ref{fig1}.
This setup can be used to realize nanoscale plasmonic isolators, a key application of nonreciprocity in photonics.\cite{fujita_2000,yu_2009b} Another approach employs plasmon resonance splitting under a DC current, see Fig.\ref{fig2}. Such a setup, which relies on current-tunable coupling to external radiation near plasmon resonance, can be used to excite and detect unidirectionally propagating SPP modes. 

Tuning carrier density and temperature further enhances the effect. 
For SPP modes constituents---carriers and long-range fields---only 
the former but not the latter are subject to Doppler effect.
Indeed, since the $1/r$ interaction is nearly instantaneous in the near-field range $k\gg\omega/c$ the corresponding motional effects are vanishingly small.  This suggests that Doppler effect 
is maximized by increasing the particle fraction and/or reducing the field fraction of the modes. 
This can be readily achieved in graphene by increasing temperature or tuning the carrier density towards charge neutrality.


The Doppler effect in graphene acquires unusual character due to the relativistic carrier dispersion in this material. As we will see, plasmonic  Doppler shift is described by unconventional quasi-relativistic transformations rather than the Galilean result\cite{Dyakonov93} $\omega'_{\vec k}=\omega_{\vec k}-\vec u\vec k$. 
These transformations, 
defined 
in $2+1$-dimensional Minkowski space-time
 via replacing the speed of light $c$ by the graphene Dirac velocity $v$, result in quasi-relativistic velocity addition, see Eqs.(\ref{eq:th_waves_velocity_addition}),(\ref{eq:dispersion_full}). 
It is instructive to draw an analogy between the plasmonic Doppler effect 
and the optical Fizeau effect (light drag by moving media) \cite{LandauLifshitz_v8}.
A relation between the two effects originates from the relativistic character of carrier dynamics in graphene. Similar to the optical Fizeau effect, the plasmonic Fizeau effect is described by Lorentz transformations 
and is therefore distinct from the conventional plasmonic Doppler effect described by Galilean transformations \cite{Dyakonov93}. 
However, the frequency shifts originating from the optical Fizeau effect are quite small,  on the order of 
$\Delta\omega/\omega\sim u/c$, with $u$ the medium velocity and $c$ the speed of light. In contrast, 
the plasmonic Fizeau effect can be much stronger than the conventional Fizeau effect, since the speed of light is replaced by graphene Fermi velocity, $v\approx c/300$, providing a few-hundred-times enhancement.

The relativistic effects are manifest 
in mode dispersion. In the collisionless regime, $\omega\ll \omega_F$, $vk\ll \omega_F$, we find
%
\be\label{eq:dispersion_parallel}
1=\frac{4\alpha v |k|\omega_F}{\gamma(\omega-uk) \sqrt{\omega^2-v^2k^2}+\omega^2-v^2k^2}
,\quad
\gamma=\frac1{\sqrt{1-\frac{u^2}{v^2}}}
,
\ee
with 
$\gamma$ the Lorentz time dilation factor, $\alpha=\frac{e^2}{\hbar v\kappa}$ the graphene fine structure constant, and $\omega_F=E_F/\hbar$. The dispersion 
features a characteristic mixture of Lorentz-invariant quantities ($\gamma$,  $\omega^2-v^2k^2$) and of non-invariant quantities ($|k|$, $\omega-uk$). The latter arises due to the Lorentz noninvariance of the $1/r$ 
interaction, see below. Linearized in $u/v$ this gives a first-order Doppler effect
\be\label{eq:delta_omega_1}
\frac{\Delta\omega}{\omega}=\lambda \frac{u}{v}
,\quad \lambda=\frac{\omega}{8\alpha \omega_F}
. 
\ee
Frequency shift is positive (negative) for plasmons propagating along (against) the flow.
The predicted value $\Delta\omega/\omega$ 
is nominally order-one when $u\sim v$, $\omega \sim \omega_F$, however it is somewhat diminished by the prefactor $1/8$. 

As mentioned above, plasmonic Doppler effect can be further enhanced by increasing temperature and/or by tuning towards neutrality, in such a way that the quasiparticle contribution overwhelms the field contribution. 
Indeed, reducing carrier density `unscreens' carrier-carrier interaction and makes it stronger, whereas higher temperature enhances collision frequency, both effects leading to fast two-particle collision rate $\gamma_{\rm e-e}$. 
Modes with frequencies $\omega\ll \gamma_{\rm e-e}$ are hydrodynamical, since carrier energies and momenta are conserved collectively rather than individually at low $\omega$, see below.
An estimate $\gamma_{\rm e-e}\approx \alpha^2 (k_{\rm B}T)^2/\max [E_F,k_{\rm B}T]$ gives frequencies under $10\,{\rm THz}$ for $E_F\approx k_{\rm B}T=300\,{\rm K}$. 
For first-order Doppler effect we find $\Delta\omega/\omega=\lambda u/v$ with
\be\label{eq:delta_omega_2}
\lambda=\lp2+4\frac{k_*}{k}\rp^{-1/2}
,\quad 
k_* =\frac{4\pi e^2  \bar n^2 }{3 \kappa \bar \epsilon}
,
\ee
where $\bar n$ and $\bar \epsilon$ are carrier number and energy densities, $\kappa$ is the substrate dielectric constant.
The quantity $k_*$ becomes small near charge neutrality and at high temperature. For short-wavelength modes, $k\gg k_*$, the value reaches $\lambda=1/\sqrt{2}\approx 0.7$. This is substantially larger than the values for the collisionless regime. 

%

The hydrodynamical regime 
can be aptly described by adopting methods from relativistic hydrodynamics.\cite{LandauLifshitz_v6} 
Energy and momentum conservation laws for weakly interacting Dirac particles can be put in a Lorentz-covariant form using a $3\times3$ stress tensor
\be\label{eq:eqs_motion_E}
\p^iT_{ij}=w_i
\ee
where $w_i$ describes energy and momentum rate of change due to long-range electric fields, see below. 
The form of the tensor $T_{ij}$ is constrained by general covariance requirements. It is convenient to express $T_{ij}$  in terms of the ``three-velocity''   $u_i=(\gamma,\gamma\vec{u}/v)$, $u_iu^i=1$ (using Minkowski $2+1$ dimensional notation).
The quantity $\vec u$ is  a function of $\vec x$ and $t$ describing local flow velocity. Using $u_i$ and the Minkowski metric, we can write\cite{LandauLifshitz_v6}
\be
T_{ij}=w' u_i u_i-p' g_{ij}
,\quad 
g_{ij}=\lp\begin{array}{ccc} 1&0&0\\0&-1&0\\0&0&-1\end{array}\rp
.
\ee
Here 
$p'$ is pressure, $w'=\epsilon'+p'$ is the thermal function, $\epsilon'$ is energy density (here prime indicates quantities in comoving frame). For massless relativistic particles, we can use the energy-pressure relation $p'=\frac12\epsilon'$ giving $w'=\frac32\epsilon'$. This relation holds for weakly interacting particles both at neutrality and away from neutrality, for zero or finite temperature. From now on we set the Dirac velocity equal unity, $v=1$, physical units to be restored later.

Below we analyze a quasi-1D flow and modes propagating along or opposite to the flow.  
We describe the system by $1+1$ dimensional space-time $(t,x)$, ignoring the dimension transverse to the flow. 
Expressing $p'$ and $w'$ through $\epsilon=\epsilon'\gamma^2$, the stress tensor is written as 
\be
T_{ij}=\lp\begin{array}{cc} w'\gamma^2-p'& w'\gamma^2 u \\ w'\gamma^2 u & w'\gamma^2 u^2 +p' \end{array}\rp
=\lp\begin{array}{cc} \epsilon\frac{2+u^2}2&\frac32\epsilon u \\ \frac32\epsilon u & \epsilon\frac{1+ 2u^2}2\end{array}\rp
\ee
(we suppressed the spatial component of $T_{ij}$ transverse to the flow and thus decoupled from longitudinal dynamics). 
%
%

We first ignore long-range electric fields, $w_i=0$, and find
collective modes by linearizing $T_{ij}$ on top of a stationary flow background
via $\epsilon=\bar \epsilon+\delta \epsilon$, $u=\bar u+\delta u$. This gives two coupled first-order differential equations for $\delta \epsilon$ and $\delta u$. 
Setting $\delta \epsilon,\delta u\sim e^{ikx-i\omega t}$ yields dispersion relation
$(2-\bar u^2)\omega^2-2\bar u \omega k+(2\bar u^2-1)k^2=0$.
Factoring this expression gives velocities for the modes propagating along and against the flow:
\be\label{eq:th_waves_velocity_addition}
(\omega-s_+ k)(\omega-s_- k)=0,\quad
s_\pm=
\frac{\bar u\pm v'}{1\pm\frac{\bar u v'}{v^2} }
,
\ee
where $v'=v/\sqrt{2}$ (we restored the Dirac velocity $v$ for clarity). 
The velocities  $s_\pm$ obey relativistic addition rules,  in accord with the above discussion.

As a consistency check, 
we consider a non-moving gas, $\bar u=0$. 
In this case, we find a wave equation $\lp 2\p_t^2-\p_x^2\rp \delta \epsilon(x,t)=0$ which describes waves propagating with the velocity $v'=v/\sqrt{2}$ as expected from relativistic hydrodynamics.\cite{LandauLifshitz_v6}
We note that Lorentz-covariant hydrodynamics described by $\p T=0$ is only valid at leading order in particle collision rate. 
Since Coulomb interactions between carriers are non-retarded, the effects such as thermal conductivity and viscosity, which arise at subleading order in the scattering rate, are not Lorentz invariant. These effect, however, will not be essential for our discussion.

A more important source of Lorentz-noninvariant contributions are long-range electric fields. 
These contributions arise at first order in interaction strength whereas the carrier-carrier scattering rate is a second-order effect. 
Since Coulomb interaction is Lorentz-noninvariant, the resulting mode 
dispersion is Lorentz nonivariant away from neutrality. However, Lorentz symmetry is restored at neutrality where collective modes are dominated by temperature dynamics and feature no charge dynamics. 

Charge dynamics can be easily incorporated in this approach. We can describe electric charge and current using Minkowski $2+1$ dimensional notation as a three-vector $j^i=(\rho,\vec j)$. Charge continuity equation, written in a Lorentz-covariant form, is
$\p_ij^i=0 $.
In the hydrodynamical regime we have
$\rho=en(\vec x,t)$, $\vec j=en(\vec x,t)\vec u(\vec x,t)$, 
where $n(\vec x,t)$ is the local carrier density and $\vec u(\vec x,t)$ is local flow velocity. 
We can also write current in a Lorentz-covariant form, $j^i=\gamma (n_0, n_0 \vec u)$, 
where $n_0$ is the carrier density in the frame comoving with the flow. 

Charge fluctuations create the long-range electric field ${\vec E}(\vec x)=-\nabla \int d^2x' V(\vec x-\vec x') n(\vec x')$, where $V(\vec x-\vec x') $ is the Coulomb potential.  This field couples to the equations of motion for energy and momentum, Eq.(\ref{eq:eqs_motion_E}), via
\be\label{eq:E-coupling_w}
w_0=\vec j\cdot {\vec E}
,\quad
w_1=en{\vec E}_x
\ee
Here the term $\vec j\cdot {\vec E}$ describes work done by the field on moving charges, whereas the term $en{\vec E}_x$ describes the Coulomb force. Particle density, which obeys the continuity equation
$\p_t n+\p_x(n u)=0$, 
becomes a dynamical variable, to be treated on equal footing with $\epsilon$ and $u$.

To describe the SPP modes in the presence of a flow, 
we linearize the equations of motion with respect to perturbations $\delta \epsilon$, $\delta u$ and $\delta n$. Taking into account that the field ${\vec E}$ arises solely due to charge fluctuations
and using 2D Fourier harmonics, we write $e{\vec E}_k=-ikV(k)\delta n_k$, where $V(k)=2\pi e^2/|k|\kappa$. This gives
\bea\nonumber
&&\lb \frac{2+\bar u^2}2\p_t+\frac32\bar u\p_x\rb \delta \epsilon +\bar \epsilon \lb \bar u\p_t+\frac32\p_x\rb\delta u=e\bar n\bar u{\vec E}
,
\\\nonumber
&&\lb \frac32\bar u\p_t+\frac{1+2\bar u^2}2\p_x\rb \delta \epsilon  + \bar \epsilon \lb \frac32\p_t+2\bar u\p_x\rb\delta u=e\bar n{\vec E}
,
\\ \label{eq:dynamics_charged}
&&\lb \p_t+\bar u\p_x\rb\delta n+\bar n\p_x\delta u=0
.
\eea
To construct plane wave solutions, $\delta \epsilon,\delta u,\delta n\sim e^{ikx-i\omega t}$, we first use the continuity equation to express density fluctuations through velocity fluctuations:
$\delta n=\frac{\bar n k}{\omega-\bar u k}\delta u$.
Plugging this in the expression for ${\vec E}_k$, we eliminate $\delta n$ and obtain two coupled linear equations for $\delta \epsilon$ and $\delta u$,
\be\label{eq:matrix_dEdu}
\lp\begin{array}{cc} \frac{2+\bar u^2}2\omega -\frac32\bar u k& \bar \epsilon\lp \bar u\omega-\frac32 k\rp-\bar u\lambda_{\omega,k} \\ \frac32\bar u\omega -\frac{1+ 2\bar u^2}2 k&\bar \epsilon\lp  \frac32\omega-2\bar u k\rp-\lambda_{\omega,k}\end{array}\rp
\lp\begin{array}{c}\delta \epsilon \\ \delta u\end{array}\rp=0
.
\ee
Here we defined the quantity $\lambda_{\omega,k}=\bar n^2V(k)\frac{k^2}{\omega-\bar u k}$ describing charge coupling.
Setting the determinant of this matrix to zero, we find dispersion relation. For a non-moving state, $\bar u=0$, this yields a relation 
$\omega^2=\frac12 v^2k^2+v^2 k_* k
$
which is identical to that found in Ref.\cite{Phan_2013}. For $\bar u\ne 0$, after some agebra we obtain 
\be\label{eq:dispersion_full}
(\omega-s_+ k)(\omega-s_-k)=\frac{1-\bar u^2}{1-\frac12\bar u^2} v^2 k_* k
\ee
with $s_\pm$ defined in Eq.(\ref{eq:th_waves_velocity_addition}).
The prefactor $\frac{1-\bar u^2}{1-\frac12\bar u^2}$ describes the weakening of charge coupling due to the flow (it equals $\frac2{\gamma^2+1}$
if expressed through Lorentz factor).
As illustrated in Fig.\ref{fig1}, the dispersion relation in Eq.(\ref{eq:dispersion_full}) becomes highly asymmetric at $u\sim v$, with one of the opposite-going modes softening and the other hardening. 
Strong nonreciprocity achieved in this regime is manifest in the order-one Doppler shift, Eq.(\ref{eq:delta_omega_2}).

Hydrodynamical modes exist for $\omega\lesssim\gamma_{\rm e-e}$. Estimating the  carrier scattering rate at neutrality as $\gamma_{\rm e-e}\approx A \alpha^2 k_{\rm B}T/\hbar$, where $A\approx 1$, for $T=300\,{\rm K}$ we obtain $\gamma_{\rm e-e}\approx 2\pi\cdot 10\,{\rm THz}$. The times $\gamma_{\rm e-e}^{-1} \approx 20 \, {\rm fs}$ are at least an order of magnitude shorter than typical disorder scattering times, enabling  weakly damped mode propagation.

Lastly we discuss the high-frequency regime $\omega\gg \gamma_{\rm e-e}$ described by  collisionless transport equations
\bea
&&\lp \p_t+\vec v_{\vec p}\nabla_{\vec x}+e {\vec E}(\vec x,t)\nabla_{\vec p}\rp n(\vec x,\vec p,t)=0
,\quad
\\
&& {\vec E}(\vec x,t)=-\nabla\int d^2x'\frac{e^2 \delta n(\vec x',t)}{|\vec x-\vec x'|}
,
\eea
%
where $\vec v_{\vec p}=\nabla_{\vec p}E(\vec p)$ is particle group velocity. Here the electric field ${\vec E}(\vec x,t)$ arises from time-varying charge fluctuations, $\delta n(\vec x,t)=N\sum_{\vec p}n(\vec x,\vec p,t)-n_0$, where $n_0$ is a compensating gate-induced background,
and $N=4$ is valley/spin degeneracy. 
In the presence of a drift,  
carriers are described by a skewed Fermi distribution
\be\label{eq:n_skewed}
n(\vec p)=\frac1{e^{\beta(E(\vec p)-\vec u\vec p-\mu)}+1},\quad E(\vec p)=v|\vec p|
,
\ee
where $\vec u$ is the drift velocity, $\mu$ is the chemical potential. 
%


We consider a weak harmonic perturbation of the uniform flow, Eq.(\ref{eq:n_skewed}). Writing $n(\vec p)+\delta n(\vec p)e^{-i\omega t+i\vec k\vec x}$ and linearizing in $\delta n$, yields dispersion relation 
\be\label{eq:1=VPi}
1=NV(k)\Pi(\vec k,\omega)
,\quad
\Pi(\vec k,\omega)=-\sum_{\vec p}\frac{\vec k\nabla_{\vec p} n(\vec p)}{\omega-\vec v_{\vec p}\vec k}
.
\ee
%
First-order Doppler effect (\ref{eq:delta_omega_1}) in the long-wavelength limit is obtained by expanding in $kv/\omega \ll 1$. 
However, since 
practically attainable drift velocities in graphene can be a substantial fraction of $v$, it is of interest to explore the full dependence on $u$ without performing any expansions. This analysis is simplified if 
Lorentz symmetry is exploited\cite{see Supplementary Information}, leading to Eq.(\ref{eq:dispersion_parallel}) discussed above.


Summing up, the plasmonic Doppler effect---dragging plasmons by a DC current flow---offers appealing ways to create nonreciprocal one-way modes in high-mobility 2DES such as graphene. The effect is strong due to carrier drift velocity high values reachabe in these materials. The magnitude and sign of nonreciprocity are fast-tunable by a DC current. These features, as well as the broad-band charachter of the plasmonic Doppler effect, make it a promising tool for achieving nonreciproicity without relying on previously explored mechanisms such as magnetic coupling, optical nonlinearity or optical pumping. Several approaches can be used to explore and uitilize the effect. Mach-Zehnder plasmonic isolators (Fig.\ref{fig1}), wherein the interference phase is controlled by a DC current through one of the arms, can be used to detect and probe nonreciprocity. Also, plasmon resonance splitting under a DC current (Fig.\ref{fig2}) can be used to excite and detect unidirectionally propagating plasmons. 

We are grateful to D. Englund, J. D. Joannopoulos, F. H. L. Koppens, M. Soljacic for useful discussions. This work was supported in part by the U. S. Army Research Laboratory and the U. S. Army Research Office through the Institute for Soldier Nanotechnologies, under contract number W911NF-13-D-0001.

\section{Supplementary Information: Quasi-relativistic Lorentz Transformations in graphene}

Here we discuss the Lorentz transformations that play a special role in our problem. 
Lorentz transformations for graphene carriers are defined 
as canonical Lorentz transformations 
in $2+1$-dimensional Minkowski space-time albeit with 
the speed of light $c$ replaced by the graphene Dirac velocity $v$. 
As a simple example, consider Lorentz boosts, defined in the usual way as 
\be\label{eq:lorentz_boost}
 t'=\gamma \lp t - \frac{u}{v^2} x\rp,
\quad  x'=\gamma (x-ut)
,\quad y'=y
\ee
Here  $\gamma = 1/\sqrt{1-u^2/v^2}$ is the Lorentz factor, and  the boost velocity $u$ is taken to be in the  $x$ direction. 
Composition of boosts with collinear velocities yields a standard velocity addition rule $u_{12}=(u_1 + u_2)/(1 +u_1 u_2/v^2)$, whereas boosts
with non-collinear velocities yield a non-commutative addition rule, described  by a more complicated expression 
\be\label{eq:u12_general}
u_{12}=\frac{\tilde{\vec u}_1 +\vec  u_2}{1 +\frac{\vec u_1 \vec u_2}{v^2}},\quad
\tilde{\vec u}_1=\vec u_{1,\parallel}+\sqrt{1-\frac{u_2^2}{v^2}}\vec u_{1,\perp}
,
\ee
where the parallel and perpendicular components of $\vec u_1$ are chosen with respect to the $\vec u_2$ direction 
(see Eq.(\ref{eq:Boltzmann_eqn_transformed_2}) and accompanying discussion). The non-commutativity is related to the fact that a composition of two boosts with non-collinear velocities does not represent as a simple boost. Instead, it is a composition of a boost with velocity $\vec u_{12}$ and a space rotation (Thomas precession).

The transformation rules for particle energy and momentum can be inferred following the standard route. We demand that the classical action $S=\int pdx-Hdt$ is invariant under Lorentz transformations, Eq.(\ref{eq:lorentz_boost}). This gives
%
\be\label{eq:lorentz_boost_Ep}
E'=\gamma(E-u p_x),
\quad p'_x=\gamma \lp p_x- \frac{u}{v^2}E\rp 
,\quad
p'_y=p_y
.
\ee
This transformation preserves the massless Dirac dispersion relation $E(p)=\pm v|\vec p|$, separately for the particle branch (plus sign) and the hole branch (minus sign).
We note that Eq.(\ref{eq:lorentz_boost_Ep})  is consistent with the Lorentz invariance of the  quantum mechanical phase in $\psi(x,t)\sim e^{(i\vec p\vec x-iEt)/\hbar}$, which provides an alternative argument for its validity.



The massless Dirac dispersion  is invariant under  Lorentz transformations, Eq.(\ref{eq:lorentz_boost_Ep}). This invariance property singles out these transformations as a symmetry of the Hamiltonian. This is similar to how Galilean transformations preserve the nonrelativistic dispersion relation $E(p)=p^2/2m$. As discussed in the main text, Lorentz symmetry has a direct impact on the Doppler effect for collective modes, which is different for a relativistic {\it vs.} a nonrelativistic carrier dispersion.

The implications of Lorentz transformations for transport in the interacting electron-hole plasma
can be clarified by analyzing the Boltzmann kinetic equation
\bea
\nonumber
(\partial_t + \vec{v}(p) \nabla_{\vec{x}}+e {\vec E}\nabla_{\vec p})n(p,x,t) = I[n(p,x,t)]
\eea
Here $n(p,x,t)$ is particle distribution function, $I[n]$ is the collision integral,  $\vec v(p)=v^2\frac{\vec p}{E(p)}$ is massless Dirac particle velocity, and ${\vec E}$ is electric field. Here $\vec v$ and $\vec p$ are collinear for the particle branch ($E(p)=v|\vec p|$) and anticollinear  for the hole branch ($E(p)=-v|\vec p|$). 
However, coupling to the long-range electric field is not Lorentz-invariant [see discussion in the main text preceding Eq.(\ref{eq:E-coupling_w})]. 
We will thus do our analysis temporarily ignoring the term $e {\vec E}\nabla_{\vec p}$ 
(the long-range field contribution can be incorporated at the last stage, see discusion in the main text). 

Under Lorentz boost,  Eq.(\ref{eq:lorentz_boost}), partial derivatives transform as
\bea
\nonumber
\p_{t}= \gamma \p_{t'} - \gamma u\p_{x'}, \quad \p_x= \gamma \p_{x'} - \gamma \frac{u}{v^2} \p_{t'}
,\quad \p_y=\p_{y'}
.
\eea
One can write the transformed kinetic equation in the new frame as
\be\label{eq:Boltzmann_eqn_transformed_1}
\lp \gamma \lp 1-\frac{uv_x}{v^2}\rp \partial_{t'} +\gamma  (v_x-u) \nabla_{x'}+v_y\p_{y'}\rp n = I(n)
,
\ee
where we used velocity components $\vec v(p)=(v_x,v_y)$. 
Pulling out the factor $\gamma\lp 1-\frac{uv_x(p)}{v^2}\rp$, we can put the result into a vector form, 
\be \label{eq:Boltzmann_eqn_transformed_2}
\gamma \lp 1-\frac{\vec{u} \vec{v}(p)}{v^2}\rp\lp \partial_{t'} + \lp \frac{\tilde {\vec v}(p)-\vec{u}}{1-\frac{\vec{u} \vec{v}(p)}{v^2}}\rp \nabla_{\vec{x}'}\rp n = I(n)
,
\ee
where we defined $\tilde {\vec v}(p)= {\vec v}_\parallel (p)+\gamma^{-1}{\vec v}_\perp (p)$, with the parallel and perpendicular components  chosen relative to the direction of $\vec u$. 

We note that the prefactor of $\nabla_{\vec x'}$ can be identified with the velocity found from the relativistic velocity addition recipe applied to $-\vec u$ and $\vec v(p)$, see Eq.(\ref{eq:u12_general}). Further, 
this velocity corresponds to particle momentum $\vec p'$  transformed according to Eq.(\ref{eq:lorentz_boost_Ep}), $\vec v(p')=v^2\frac{\vec p'}{E'(p')}$. This gives
\be 
(\partial_{t'} + \vec v(p') \nabla_{\vec x'})n = \frac{E(p)}{E'(p')}I(n)
,
\ee
where we used the identity $\gamma \lp 1-\frac{\vec{u} \vec{v}(p)}{v^2}\rp=E'(p')/E(p)$, giving a factor $E/E'$ before $I(n)$. Along with the the velocity addition recipe, the factor $E/E'$ routinely appears in relativistic physics, where it describes the change in collision frequency upon transformation to a moving frame. 

For a Lorentz invariant interaction, the factor $E/E'$ could be absorbed in $I(n)$. In our case, however, carrier scattering is governed by the Coulomb interaction which is Lorentz non-invariant.  
We see that
the collision term, because of the factor $E/E'$, is not Lorentz invariant. However, the factor $E/E'$ has no impact on the form of the resulting transport equations at leading order in small $k$ and $\omega$. Indeed, particle distribution  in the hydrodynamical regime obeys local equilibrium, satisfying the zero-mode condition,\cite{Phan_2013} $I(n)=0$. We therefore conclude that, despite the non-invariant collision term, the transport equations for our problem are Lorentz invariant in the limit of a high particle collision rate, which defines the hydrodynamical regime. 

\section{Polarization operator in the presence of a flow}

Here we analyze the polarization operator $\Pi(\vec k,\omega)$ which governs the dependence of plasmon dispersion on the flow velocity $\vec u$, see Eq.(\ref{eq:1=VPi}). In this analysis it will be convenient to use a quasi-relativistic Lorentz transformation for $E(\vec p)$ and $\vec p$. To exhibit the quasirelativistic aspects of the problem, and thereby to simplify our analysis, we introduce an energy variable which is independent of momentum variable. This is done by rewriting the expression for $\nabla_{\vec p} n(\vec p)$
by inserting an integral over a new independent variable $0<\epsilon<\infty$:
\be 
\nabla_{\vec p} n(\vec p) = 
\int d\epsilon \delta(\epsilon-E(\vec p)) \lp \vec v_{\vec p}-\vec u\rp \frac{\p n_{\epsilon, \vec p}}{\p\epsilon}
\ee
where $n_{\epsilon, \vec p}=1/(e^{\beta(\epsilon-\vec u\vec p-\mu)}+1)$. Plugging it in the expression for $\Pi(\vec k,\omega)$ 
and writing $\vec v_{\vec p}=v^2\vec p/\epsilon$, we obtain
\be \label{eq:Pi2}
\Pi(\vec k,\omega)=-\sum_{\vec p}\int  d\epsilon \delta(\epsilon-v|\vec p|)\frac{v^2\vec k\vec p-\vec k\vec u\epsilon }{\omega\epsilon-v^2\vec k\vec p} \frac{\p n_{\epsilon, \vec p}}{\p\epsilon}
,
\ee
The above expression can be simplified by going to new variables defined according to Lorentz transformation rules. Without loss of generality we choose $\vec u$ to be in the $x$ direction. We define
\be
\epsilon'=\gamma (\epsilon-up_x),\quad p'_x=\gamma \lp p_x-\frac{u}{v^2}\epsilon\rp
,\quad
p'_y=p_y
\ee
with $\gamma=(1-u^2/v^2)^{-1/2}$, which corresponds to a Lorentz boost to the frame comoving with the flow. In the new frame, particle momentum distribution becomes isotropic. In agreement with this intuition, we have 
\be
\frac{\p n_{\epsilon, \vec p}}{\p\epsilon}=\gamma \frac{\p \tilde n(\epsilon')}{\p\epsilon'}
,\quad
\tilde n(\epsilon)=\frac1{e^{\tilde \beta(\epsilon-\tilde\mu)}+1}
,
\ee
where $\tilde \beta=\beta/\gamma$, $\tilde\mu=\mu\gamma$. 

To perform Lorentz transformation of variables under the integral, Eq.(\ref{eq:Pi2}), we note that $\epsilon^2-v^2\vec p^2$ is Lorentz invariant, and therefore $\delta(\epsilon-v|\vec p|)=\frac{\epsilon'}{\epsilon}\delta(\epsilon'-v|\vec p'|)$. 
Other quantities in Eq.(\ref{eq:Pi2}) transform as
\be
v^2\vec k\vec p-\vec k\vec u\epsilon=v^2\vec k_\gamma\vec p'
,\quad
\omega\epsilon-v^2\vec k\vec p=\omega'\epsilon'-v^2\vec k'\vec p'
\ee
where $\vec k_\gamma=(k_x,\gamma k_y)$. 
Here we defined primed (Lorentz-transformed) frequency and wavevector
\be
\omega'=\gamma (\omega-uk_x),\quad k'_x=\gamma \lp k_x-\frac{u}{v^2}\omega\rp
,\quad
k'_y=k_y
\ee
Combining these results, we find
\be 
\Pi(\vec k,\omega)=-\sum_{\vec p'}\int  d\epsilon' \frac{\epsilon}{\epsilon'}\delta(\epsilon'-E(\vec p'))\frac{v^2\vec k_\gamma \vec p'}{\omega'\epsilon'-v^2\vec k'\vec p'} \frac{\p \tilde n(\epsilon')}{\p\epsilon'}
\ee
Next,  removing the integral over $\epsilon'$ and the delta function, and setting $\epsilon'=E(\vec p')$, gives
\be 
\Pi(\vec k,\omega)=-\sum_{\vec p'}\frac{\epsilon}{\epsilon'}\frac{v^2\vec k_\gamma \vec p'}{\omega'\epsilon'-v^2\vec k'\vec p'} \frac{\p \tilde n(\epsilon')}{\p\epsilon'}
\ee
Lastly, expressing $\epsilon$ through  $\epsilon'$, $\vec p'$ and introducing the velocity $\vec v'=v^2\vec p'/\epsilon'$, we obtain an expression 
\be 
\Pi(\vec k,\omega)=-\sum_{\vec p'}\gamma \lp 1+\frac{\vec u\vec v'}{v^2}\rp \frac{\vec k_\gamma \vec v'}{\omega'-\vec k'\vec v'} \frac{\p \tilde n(\epsilon')}{\p\epsilon'}
\ee
In the above expression, the last term depends on $|\vec p'|$ and has no angle dependence, whereas other terms depend on the direction of $\vec v'$ but not on  $|\vec p'|$. Using this property to separate angular and radial integration over $\vec p'$, we write the result as
\be \label{eq:Pi}
\Pi(\vec k,\omega)=-\left\la \lp 1+\frac{\vec u\vec v'}{v^2}\rp \frac{\vec k_\gamma \vec v'}{\omega'-\vec k'\vec v'} \right\ra_\theta \gamma\int d\epsilon' N(\epsilon')\frac{\p \tilde n(\epsilon')}{\p\epsilon'}
\ee
where $N(\epsilon)$ is the density of states in graphene. 
Here the brackets stand for angle-averaging, $\la ...\ra_\theta=\oint ...\frac{d\theta}{2\pi}$, where $\theta$ is the angle between $\vec v'$ and $\vec u$. 

We now proceed to analyze the quantity $\la ...\ra_\theta$ which contains the essential dependence on the drift velocity $\vec u$. First, we consider the case of $\vec k\parallel\vec u$. We have
\be
\la ...\ra_\theta=\oint \frac{d\theta}{2\pi}\lp 1+\frac{u}{v}\cos\theta\rp \frac{kv\cos\theta}{\omega'- k'v\cos\theta}
\ee
Evaluating the integral, we find
\be
\la ...\ra_\theta=\frac{k}{k'}\lp 1+\frac{u\omega'}{v^2k'}\rp \lp-1+\frac{\omega'}{\sqrt{{\omega'}^2-v^2{k'}^2}}\rp
\ee
Taking into account the relation $k'+u\omega'/v^2=k/\gamma$, we rewrite the result as
\be
\la ...\ra_\theta=\frac{k^2v^2}{\gamma\lp\omega' \sqrt{{\omega'}^2-v^2{k'}^2}+{\omega'}^2-v^2{k'}^2\rp}
\ee
Lastly, using the identities ${\omega'}^2-v^2{k'}^2=\omega^2-v^2k^2$, $\omega'=\gamma(\omega-uk)$, we find
\be
\la ...\ra_\theta=\frac{k^2v^2}{\gamma\lp\gamma(\omega-uk) \sqrt{\omega^2-v^2k^2}+\omega^2-v^2k^2\rp}
\ee
Plugging this result in Eq.(\ref{eq:Pi}) and then in Eq.(\ref{eq:1=VPi}) we can write plasmon dispersion as
\be
1=\frac{2\pi e^2 |k|v^2 a}{\gamma(\omega-uk) \sqrt{\omega^2-v^2k^2}+\omega^2-v^2k^2}
.
\ee
Here $a=-\int d\epsilon' N(\epsilon')\frac{\p \tilde n(\epsilon')}{\p\epsilon'}$ where $N(\epsilon)$ is the density of states. For an electron system at degeneracy, $k_{\rm B}T\ll\mu$, we can approximate the density of states by its value at the Fermi level, giving $a\approx N(\epsilon=\mu\gamma)$.


Using the standard expression for the density of states,
$
N(\epsilon)=\frac{2\pi N p}{(2\pi\hbar)^2v}=\frac{2\epsilon}{\pi\hbar^2v^2}
$
we can put our result in the form given in Eq.(\ref{eq:dispersion_parallel}). 
This expression can be simplified in the long-wavelength limit $vk\ll\omega$, giving $\gamma (\omega-uk) \omega+\omega^2=4\alpha\omega_F v|k|$. 
Expanding in a small nonzero $u$ we have
\be
\omega^2\approx 2\alpha\omega_F v|k|+\frac12 uk\omega\approx 2\alpha\omega_F v|k|\lp 1+\frac{uk}{2\omega}\rp
\ee
giving a Doppler shift of $\frac{\Delta\omega}{\omega}=\frac{u\omega}{8v\alpha\omega_F}$ which matches Eq.(\ref{eq:delta_omega_1}).


%
We also quote the dispersion for $u=0$:
\be
\omega = \frac{2\lambda |k| + v^2k^2}{\sqrt{4\lambda |k| + v^2 k^2}}
\ee
In the long-wavelength limit it reproduces the standard square-root plasmon dispersion, $\omega\sim k^{1/2}$. 


\end{document}